\documentclass[a4paper,11pt]{article}
\pdfoutput=1 
\usepackage{jheppub} 
\usepackage{epsfig}
\usepackage{caption}
\usepackage{subcaption}
\usepackage{color,soul,bm}
\usepackage{float}
\usepackage{amsfonts}
\usepackage{url}
\usepackage{slashed}
\usepackage{accents}
\usepackage{lipsum}
\usepackage{mathtools}
\usepackage{physics}

\usepackage[normalem]{ulem}

\hypersetup{
  bookmarks=true,         
  unicode=false,          
  pdftoolbar=true,        
 pdfmenubar=true,        
 pdffitwindow=true,     
 pdfstartview={FitH},    
 pdfsubject={Dark Matter},   
 pdfnewwindow=true,      
 pdfcreator={RevTeX},
 colorlinks=true,       
 linkcolor=red,          
 citecolor=blue,        
 filecolor=black,      
 urlcolor=blue,           
  }

\title{Improved Treatment of Bosonic Dark Matter Dynamics in Neutron Stars: Consequences and Constraints}
\author[a]{Koushik Dutta,}
\author[a]{Deep Ghosh}
\author[a]{and Biswarup Mukhopadhyaya}
\affiliation[a]{Department of Physical Sciences, Indian Institute of Science Education and Research (IISER) Kolkata, Campus Road, Mohanpur, West Bengal 741246.}
\emailAdd{koushik@iiserkol.ac.in}
\emailAdd{matrideb1@gmail.com}
\emailAdd{biswarup@iiserkol.ac.in}
\abstract{
It is conceivable that a bosonic dark matter (DM) with non-gravitational interactions with SM particles will be accumulated at the center of a neutron star (NS) and can lead to black hole formation. \textit{In contrast to previous works with a fixed NS temperature, we dynamically determine the formation of Bose-Einstein condensate (BEC) for a given  set of DM parameters, namely the DM-neutron scattering cross-section ($\sigma_{\chi n}$), the thermal average of DM annihilation cross-section ($\expval{\sigma v}$) and the DM mass ($m_\chi$)}. For both non-annihilating and annihilating DM with $\expval{\sigma v} \lesssim 10^{-26}{~\rm cm^3~ s^{-1}}$, the BEC forms for $m_\chi \lesssim 10$ TeV. In case of non-annihilating DM, observations of old NS allows $\sigma_{\chi n}\lesssim 10^{-52}~{\rm cm^2}$ for $10 {~\rm MeV} \leq m_{\chi} \lesssim 10 {~\rm GeV}$ (with BEC) and  $\sigma_{\chi n}\lesssim 10^{-47}~{\rm cm^2}$ for $5 {~\rm TeV} \lesssim m_\chi \lesssim 30 {~\rm PeV} $ (without BEC). This analysis shows that the electroweak mass window, $10 {~\rm GeV} \lesssim m_\chi \lesssim 5  {~\rm TeV}$ is essentially     unconstrained by NS observations and therefore is subject only to direct detection experiments. In the annihilating DM scenario, the exclusion limits on DM parameters become weaker and even vanish for typical WIMP annihilation cross-section. However, the late-time heating of the NS enables us to probe the region with $\sigma_{\chi n}\gtrsim 10^{-47}~{\rm cm^2}$, using the James Webb Space Telescope in the foreseeable future. When our results are viewed in the context of indirect searches of DM, it provides a lower limit on the $\expval{\sigma v}$, which is sensitive to the DM thermal state.
}

\begin{document} 
\maketitle
\flushbottom  
\section{Introduction and Summary}

With the advent of new age ground-based and space-based telescopes, neutron stars (NS) are going to be promising laboratories for detecting non-gravitational interactions of particle dark matter (DM), over and above to direct and indirect detection experiments. In particular, direct detection experiments (e.g. XENON1T \cite{XENON:2017vdw}, LZ\cite{LUX:2013afz}, CRESST-III\cite{CRESST:2017cdd}) are instrumental in probing the DM-nucleon elastic scattering cross-section ($\sigma_{\chi n}$), whereas the thermally averaged cross-section ($\expval{\sigma v}$) of DM annihilation to visible Standard Model (SM) states is relevant to indirect searches through various telescopes (e.g. Fermi-LAT\cite{Fermi-LAT:2009ihh}, AMS-02\cite{AMS:2016oqu}) and dedicated neutrino detectors (e.g. SK\cite{Super-Kamiokande:2017yvm}, DUNE\cite{DUNE:2020ypp}, IceCube\cite{IceCube:2011ucd}). In the context of canonical WIMP scenarios, $\expval{\sigma v}$ also determines the observed DM relic density, comprised of particles with masses ($m_\chi$) within GeV-TeV range. Here, we propose to revisit the constraints on these DM parameters ($\sigma_{\chi n}$, $\expval{\sigma v}$, $m_\chi$) in the context of neutron stars, as outlined below.  

The first hint of probing DM interactions in NSs comes from the ``missing pulsar problem" \cite{Pfahl:2003tf,Macquart:2010vf,Dexter:2013xga}, which questions the null observation old millisecond pulsars near the center of the Milky Way galaxy. In Ref.\cite{Bramante:2014zca}, it has been pointed out that DM acquisition inside the NS can lead to the black hole (BH) formation and eventual destruction of the host star, thereby explaining the missing pulsar population. In this context, the observation of an old NS with typical age about a Gyr near the galactic center \cite{GC,Bower:2013tva} or any other DM-rich environments (like globular clusters \cite{Camilo:2005aa,Padmanabh:2024bsz,Brown:2018pwq,Reynoso-Cordova:2022ojo}) can constrain the DM capture rate. It is evident that the constraint is dependent on the age of the NS as well as the DM density around it. In particular, the older the NS and the denser the local DM halo, the stronger the constraint on the DM capture rate becomes. Now, the DM capture takes place only when the gravitationally in-falling DM particles dissipate their kinetic energy via elastic scatterings with neutrons and eventually get trapped inside the NS. Consequently, the DM-neutron scattering cross-section\footnote{From now on, $\sigma_{\chi n}$ represents the DM-neutron scattering cross-section.} ($\sigma_{\chi n}$) get constrained from the DM capture rate for a given DM mass. In fact, the constraint on the $\sigma_{\chi n}$ is expected to be stronger and applicable for a wider mass range than in the terrestrial direct detection experiments due to higher density of neutrons in NSs \cite{deLavallaz:2010wp,Raj:2017wrv,Maity:2021fxw,Linden:2024uph}.

Now, with non-zero annihilation rate of DM particles inside the NS, the upper bound on $\sigma_{\chi n}$ weakens. Moreover it can become completely unconstrained from the mere existence of old NS in presence of \textit{sufficiently} large DM annihilation cross-section. However, the DM annihilation to SM states confined within the NS modifies the standard evolution of NS cooling \cite{Yakovlev:2003qy}, that results in the late-time heating of an old NS. Thermal emissions from the surface of an old NS, associated with this late-time heating can be detectable in current and future probes, such as the James Webb Space Telescope (JWST)\cite{JWST}, Thirty Meter Telescope (TMT)\cite{TMT} or Extremely Large Telescope (ELT)\cite{ELT} with sensitivities in the visible and infra-red frequencies. Such a detection can again constrain the $\sigma_{\chi n}$ which decides the amount of heating due to DM annihilation in the \textit{capture-annihilation equilibrium} condition \cite{Kouvaris:2007ay}.

In the light of the above discussion, we note that NS observations can constrain $\sigma_{\chi n}$ for both non-annihilating and annihilating DM scenarios. However, the thermal properties of captured DM particles become important in determining constraints on the DM parameters. For thermalized DM particles accumulated in the interior of a NS form a dark core with a radius given by the NS temperature and the DM mass. In case of the dark core collapse, thermalization of DM particles is a sufficient condition \cite{Bell:2013xk}. For bosonic DM particles, the thermalized dark core consists of either a Bose-Einstein condensate (BEC) or a non-BEC state, which determine the BH formation subsequently. In fact, the BH formation associated with a BEC state, requires less number of DM particles than in a non-BEC state. This is due to the fact that for a non-BEC state, the collapse is somewhat hindered due to pressure support, stemming from large occupation number in high momentum states, unlike the BEC state which is primarily associated with low momenta. Consequently, the constraint on the $\sigma_{\chi n}$ for a given DM mass would be quite different in these two cases. Noticeably, the thermalization of DM particles is established  via the same DM-neutron interaction, leading to significant observational impacts \cite{Garani:2020wge,Bell:2023ysh}.
 
In the present study, we dynamically determine (in terms of parameters namely $\sigma_{\chi n}$, $\expval{\sigma v}$, $m_\chi$) the thermal state of bosonic DM particles inside a NS, which is not considered in most of the previous studies \cite{McDermott:2011jp,Bramante:2013hn,Garani:2018kkd,Bhattacharya:2023stq,Lu:2024kiz,Liu:2024qbe,Ray:2023auh}, as per our knowledge. In particular, we at first focus on the distinction between a BEC and a non-BEC state in the DM parameter space. Then we follow its effect on possible BH formation and the late-time heating for annihilating DM scenarios. Our main findings are summarized below.
\begin{enumerate}
\item For a given set of DM parameters, we uniquely determine the BEC state forming out of accumulated DM particles inside a NS, residing in the DM environment with density of $10^3 {~\rm GeV~ cm^{-3}}$. \textit{This is done dynamically by simultaneously solving the evolution equations of the NS temperature and the DM yield - which is a novel study. In particular, the NS temperature at which BEC forms, is completely fixed by the DM parameters, unlike most of the earlier works in which the temperature comes as an input}. We additionally check the thermalization of DM particles with evolving NS temperature, necessary for subsequent BEC and BH formation. The BEC formation is determined by comparing the the ambient NS temperature and the critical temperature for BEC formation, both of which are set by the DM parameters.  

\item In both annihilating and non-annihilating DM scenarios, we find that the BEC state forms within DM masses, $10 {~\rm MeV} \leq m_\chi \lesssim 10 ~{\rm TeV}$ for $\sigma_{\chi n}$ values as shown in Fig.\ref{fig:scan} and Fig.\ref{fig:scan1} (blue solid line surrounded region).  The lower bound on the DM mass comes from the fact that thermal DM particles with $m_\chi < 10$ MeV would escape from the NS due to their higher thermal velocities than the escape velocity of the star. The upper bound on the DM mass in the non-annihilating DM scenario comes from the fact that the DM number required for the BEC formation is sufficient to collapse the dark core due to self-gravitation. In case of annihilating DM, the bound comes from an interplay between capture and annihilation processes. In particular, one needs higher DM capture rate to form a BEC for $m_\chi>10$ TeV, consequently it leads to greater heating of the NS due to annihilation. Most of the surviving DM particles acquire higher momentum states corresponding to the higher temperature, thereby hindering the BEC formation at the high mass end. 
\item For a BEC state, constraints on the $\sigma_{\chi n}$ derived from the observation of $10$ Gyr old NS are stronger than in the non-BEC state owing to larger DM particles requiring for BH formation in the latter case. In the non-annihilating DM scenario, masses within $10 ~{\rm MeV} \lesssim m_\chi \lesssim 10$ GeV, are disallowed for $\sigma_{\chi n}$ values as shown in Fig.\ref{fig:scan} (red shaded region) for the BEC state. For a non BEC state, the BH formation due to self-gravitation constrains DM masses $10 ~{\rm TeV} \lesssim m_\chi \lesssim 30 ~{\rm PeV}$ as shown in Fig.\ref{fig:scan} (green shaded region). We note that the constraint on $\sigma_{\chi n}$ for the non-BEC state is found to be weaker significantly in our dynamical analysis than that of the Refs.\cite{McDermott:2011jp,Garani:2018kkd}. For detailed discussion on apparent differences see Sec.\ref{sec:sec3}.  

\item In the non-annihilating DM scenario, the regions in the DM parameter space with $10 ~{\rm GeV} \lesssim m_\chi \lesssim 5 ~{\rm TeV} $ and $m_\chi \gtrsim 30 ~{\rm PeV}$ are not constrained from the existence of an old NS, due to BH evaporation through Hawking radiation. Therefore, the direct detection (DD) experiments remain important to probe the DM-neutron scattering cross-section for DM masses within $10 ~{\rm GeV} \lesssim m_\chi \lesssim 5 ~{\rm TeV} $. However, for sub-GeV DM masses ($10 ~{\rm MeV} \lesssim m_\chi \lesssim 10 ~{\rm GeV}$), the NS observations can be complementary to the DD experiments.     

\item In the annihilating DM scenarios, the upper bounds on the $\sigma_{\chi n}$ for BEC and non-BEC states weaken due to depletion of DM particles, signifying the survival of an old NS. We find that for canonical WIMP scenario with $\expval{\sigma v}\approx 10^{-26}{~\rm cm^3~ s^{-1}}$, there is no constraint in the $\sigma_{\chi n}-m_\chi$ plane from the existence of a $10$ Gyr old NS. However, the DM annihilation-induced late-time heating of NS has been exploited to show the parameter space for $\sigma_{\chi n}$, which can be probed by the JWST in the foreseeable future.     

\item Finally, we show the effect of annihilation by varying $\expval{\sigma v}$ over a wide range keeping $\sigma_{\chi n}$ fixed at its geometric limit. We find that  BH never forms for $\expval{\sigma v}\gtrsim 10^{-39} {~\rm cm^3~ s^{-1}}$ in the BEC state,  thus signifying a lower bound on the $\expval{\sigma v}$. For the non-BEC state the lower bound on the $\expval{\sigma v}$ strengthens with increasing DM mass, unlike the former case in which the bound is somewhat independent of the DM mass. Consequently, these bounds provide a \textit{floor} over a wide range of DM masses ($m_\chi \lesssim 10$ GeV and $m_\chi \gtrsim$ a few TeV, for details see Fig.\ref{fig:sigmav}) for indirect detection experiments, provided $\sigma_{\chi n}$ is determined. 
\end{enumerate}
The organization of this article is as follows: In Sec.\ref{sec:sec2} we discuss the setup for dynamical evolution of the NS temperature and the DM yield to discriminate between a BEC and a non-BEC state, subsequently checking BH formation in both the cases. We present our new results in detail for both annihilating and non-annihilating DM scenarios in Sec.\ref{sec:sec3}. Finally we hint the possibility of distinguishing between a BEC and a non-BEC state in terms of NS parameters in Sec.\ref{sec:sec4}.   
\section{BEC formation by dark matter and stability of the neutron star}     
\label{sec:sec2}
In this section, we take a brisk tour of DM dynamics inside the NS, particularly the interplay between capture and annihilation processes. We set up the relevant equations from which the distinction between a BEC state and a non-BEC state becomes apparent in terms of DM parameters. In addition, we also discuss various caveats of dark core collapse, leading to transmuting a NS into a BH.
\subsection{Dynamical setup }
Formation of BEC state primarily depends on the DM capture rate, the ambient temperature of the NS and the DM annihilation rate. The deposition of DM particles into an isolated NS is facilitated by strong gravitational potential of the NS. However, the in-falling DM particles can not be trapped inside a NS without any interaction between DM and neutrons present in the NS. This stems from the fact that DM particles acquire the escape velocity of the host star, thereby escape from the interior of the star. Therefore, the capture rate depends on the scattering between DM and neutrons, given the local DM density ($\rho_\chi$) as follows\footnote{For more accurate calculation of DM capture rate, including effects, like - relativistic energy transfer, Pauli blocking and multiple scatterings see Refs.\cite{Bell:2020jou,Bramante:2017xlb,Dasgupta:2019juq}.}\cite{Kouvaris:2007ay,Bell:2013xk} : 
\begin{align}
C_c&=\sqrt{6\pi}\left(\frac{\rho_\chi}{m_\chi v_\chi}\right)\left(\frac{2GM_{NS}R_{NS}}{1-\frac{2GM_{NS}}{R_{NS}}}\right)f,
\label{eq:cap}
\end{align} 
where $f={\rm Min}[1,\sigma_{\chi n}/\sigma_g]$ and $\sigma_g$ represents the geometric cross-section of the DM-neutron scattering. $v_\chi$ is the average velocity of DM particles in the galactic halo, taken to be $220 ~{\rm km/s}$. Evidently, the capture rate depends on the radius ($R_{NS}$) and the mass ($M_{NS}$) of NS, which decide its gravitational potential. For our subsequent studies, we take $R_{NS}= 10.6$ km, $M_{NS}=1.4 M_{\odot}$, $M_{\odot}$ being the solar mass. 

In addition to NS parameters, $\sigma_g$ depends on the momentum transfer of the DM-neutron collision, stemming from degenerate neutrons. It brings about its dependence on the DM mass in the following way \cite{Bell:2013xk}, 
\begin{align}
\sigma_g \approx \frac{\pi R^2_{NS}}{N_n} \left({\rm Min} \left[\frac{m_\chi}{0.2 {~\rm GeV}},1 \right]\right)^{-1},
\label{eq:geom}
\end{align} 
where the number of neutrons in the NS is denoted by $N_n= M_{NS}/m_n$ and $m_n$ is the neutron mass. The above equation indicates that for $ m_\chi<0.2$ GeV, only a fraction of neutrons are accessible for the scattering process. In particular, the fraction of total neutrons participate in scatterings with DM particles having $m_\chi << m_n$ is estimated as $3 m_\chi/p_F$, where $p_F$ is the Fermi momentum of neutrons, given by $p_F\sim 0.5$ GeV \cite{Bell:2013xk}. 

The DM-neutron interactions bring about the thermalization of DM particles at the NS core temperature $T_{NS}$. DM particles with initial velocities roughly around the escape velocity of the NS lose kinetic energy via these collisions and eventually acquire thermal velocity distribution, corresponding to $T_{NS}$. The thermalization time for captured DM particles is estimated as follows \cite{McDermott:2011jp,Bramante:2013hn} : 

\begin{align}
\label{therm}
t_{th}= 
\begin{cases}
   0.054 {~\rm years} \left(\frac{m_\chi}{100~{\rm GeV}}\right)^2 \left(\frac{2.1\times 10^{-45} {\rm cm^2}}{\sigma_{\chi n}}\right)\left(\frac{10^{-5} ~{\rm MeV}}{T_{NS}}\right),~ {\rm for} ~~m_{\chi} \gtrsim 1 ~{\rm GeV},\\
       7.7\times 10^{-5} {~\rm years} \left(\frac{0.1~{\rm GeV}}{m_\chi}\right) \left(\frac{2.1\times 10^{-45} {\rm cm^2}}{\sigma_{\chi n}}\right)\left(\frac{10^{-5} ~{\rm MeV}}{T_{NS}}\right),~ {\rm for} ~~m_{\chi} \lesssim 1~ {\rm GeV}.
\end{cases}
\end{align}
The difference in thermalization time around $m_\chi \sim 1$ GeV, originates from the difference in the momentum transfer ($\delta p$) during the elastic scatterings, depending on the DM mass as $\delta p \sim (m_\chi m_n)/(m_\chi+m_n)$. For more detailed analysis on thermalization see Refs.\cite{Bertoni:2013bsa,Bell:2023ysh,Garani:2018kkd}. Thermalized DM particles form a dark core with thermal radius $r_{th}$ given by \cite{Bell:2023ysh},
\begin{align}
r_{th} = \sqrt{\frac{3T_{NS}}{2\pi G \rho_c m_\chi}}= 26~ {\rm km} \left (\frac{T_{NS}}{m_\chi}\right)^{1/2}, 
\end{align} 
where $\rho_c = 5\times 10^{38} ~ {\rm GeV~ cm^{-3}}$, being the typical baryon density of a NS core. This stems from the balance between the gravitational force acted on DM particles due to baryon density within radius $r_{th}$ and the pressure due to thermal motions of DM particles with an approximate average velocity of $\sqrt{T_{NS}/m_\chi}$.  

A population of thermalized DM particles inside the NS can be diminished due to evaporation effect \cite{Garani:2021feo}, stemming from the thermal velocities of DM particles, above the escape velocity of the star. However, for our subsequent analyses, we take DM masses, $m_\chi \geq 10~$ MeV, for which the evaporation effect is negligible throughout the evolution of a NS with initial temperature, $T^i_{NS}=1~$MeV. This can be realized by restricting the thermal radius of the dark core as,
\begin{align}
r_{th}\lesssim R_{NS} \implies m_{\chi}/T_{NS} \gtrsim 10, ~~\text{where}, R_{NS}=10.6 ~{\rm km}. 
\label{eq:eva}
\end{align}

Once the thermalization is ensured, the BEC state can form when the temperature of the dark core ($T_\chi=T_{NS}$) falls below its critical temperature ($T_c$) given by,
\begin{align}
T_c = \frac{2\pi}{m_\chi}\left(\frac{3 N_\chi}{4\pi r^3_{th} ~\zeta(3/2)}\right)^{2/3}.
\label{eq:crit}
\end{align} 
$T_c$ determines the transition of bosons into a condensed state in which most of the particles reside in the ground state. It is apparent that $T_c$ is a dynamical quantity as it depends on the ambient number ($N_\chi$) of DM particles as well as the core temperature of a NS via its dependence through $r_{th}$. Therefore, we exploit the NS temperature evolution in presence of DM capture events to determine the epoch of BEC formation within the lifetime of a  NS. In fact, the BEC formation is completely determined by the DM parameters, namely $\sigma_{\chi n}$, $\expval{\sigma v}$ and $m_\chi$.

In general, $T_{NS}$ is a function of both space and time coordinates. We consider the evolution after the thermal relaxation phase of NS, for which $T_{NS}$ becomes only a function of time \cite{Yakovlev:2000jp}. The temperature evolution of a NS in presence of the DM capture and its possible annihilation is given by 
\begin{align}
\frac{dT_{NS}}{dt} = -\frac{\epsilon_\nu + \epsilon_\gamma - \epsilon_\chi}{c_V},
\label{eq:temp}
\end{align}
where $\epsilon_\nu$, $\epsilon_\gamma$ are standard emissivities coming from neutrino ( dominantly via mURCA process \cite{Shapiro:1983du}) and photon channels\footnote{Recently, it has been pointed out in Ref.\cite{Chakraborty:2023wgl} that there is dominant neutrino contribution ($n \gamma \rightarrow n \nu \bar{\nu}$) comes from the \textit{Wess-Zumino-Witten} term within SM for $T_{NS}\gtrsim 1$ MeV.} respectively, while $c_V$ is the total specific heat, having contribution from the dominant constituents (i.e. neutron, proton, electron) of a NS. These quantities are given as follows \cite{Kouvaris:2007ay,Shapiro:1983du,Gudmun1,Gudmun2,Page:2004fy}:
\begin{align}
\epsilon_\nu &= 3.3 \times 10^{-15} ~{\rm MeV^4~ year^{-1}} \left(\frac{T_{NS}}{10^{-3} {~\rm MeV}}\right)^8 ,\nonumber\\
\epsilon_\gamma &= 2.8 \times 10^{-5} ~{\rm MeV^4~ year^{-1}} \left(\frac{T_{NS}}{10^{-2} {~\rm MeV}}\right)^{2.2} ,\nonumber\\
c_V &= \frac{T_{NS}}{3}\sum_{i=n,p,e}p_{F,i}\sqrt{p^2_{F,i}+m^2_i}.
\end{align}
Here $p_{F,i}$ is the Fermi momentum of the $i^{th}$ particle species, dependent on its number density inside the NS. $\epsilon_\chi$ is the DM counterpart of the standard emissivities given by
\begin{align}
\epsilon_\chi =  \frac{m_\chi C_a}{\frac{4}{3}\pi R^3_{NS}} N^2_\chi,
\label{eq:demi}
\end{align} 
where $C_a$ signifies the annihilation rate of thermalized DM particles within a volume having a radius of $r_{th}$, and the expression for $C_a$ is given by $C_a=\frac{\expval{\sigma v}}{ (2\pi)^{3/2} r^3_{th}}$. 

A non-vanishing dark emissivity tends to increase $T_{NS}$, if the annihilation products lose energy inside the star via interactions primarily with electrons present at the NS crust. Hence, the temperature evolution in presence of DM annihilation should be augmented with the evolution of the DM number as the following.
\begin{align}
\frac{dN_\chi}{dt} = C_c-C_a N^2_\chi. 
\label{eq:number}
\end{align}   
To note, Eq.\ref{eq:temp} and Eq.\ref{eq:number} become coupled via the dark emissivity term, therefore the DM dynamics impacts the NS temperature evolution. In the \textit{capture-annihilation equilibrium} condition (i.e. $dN_\chi/dt=0$), the dark emissivity is determined by the capture rate only, i.e. 
\begin{align}
\epsilon_\chi = \frac{m_\chi C_c}{\frac{4}{3}\pi R^3_{NS}}.
\label{eq:dcap}
\end{align}
For an old NS of age $\gtrsim 10^8$ years, when the standard cooling process become negligible, the thermal radiation produced due to the energy deposition by DM annihilation products leads to the late-time heating of the NS. The emissivity of the dark thermal radiation from the surface of the NS can be quantified using Stefan-Boltzmann equation as,
\begin{align}
\epsilon_\chi= \frac{3 \sigma}{ R_{NS}} T^4_{sur}, 
\label{eq:drad}
\end{align}
where $\sigma=5.67\times 10^{-5}~{\rm erg~ cm^{-2}~s^{-1}~ K^{-4}}$. $T_{sur}$ is the surface temperature of the NS, which is related to the  the NS core temperature as the following \cite{Gudmun1,Gudmun2}.
\begin{align}
T_{sur} = 8.7\times 10^{-5} ~{\rm MeV}~ \left(\frac{T_{NS}}{10^{-2} {~\rm MeV}}\right)^{0.55}
\label{eq:stemp}
\end{align}
In the capture-annihilation equilibrium, the dark emissivity becomes independent of the ambient NS temperature (apparent from Eq.\ref{eq:dcap}), thereby $T_{NS}$ freezes out at a constant value, $T^f_{NS}$, signifying the late-time heating due to DM annihilation\footnote{Apart from DM annihilations, there can other sources for late-time heating of a NS. For such examples see Refs.\cite{Baryakhtar:2017dbj,Fujiwara:2023hlj,Raj:2024kjq,Ema:2024wqr} and references therein.}.
\begin{align}
T^f_{NS} \approx 1.8\times 10^{-5} {~\rm MeV}~\left(\frac{m_\chi}{{10~\rm GeV}} ~\frac{C_c}{8\times 10^{35} {~\rm year^{-1}}}\right)^{0.45}.
\label{eq:tempf}
\end{align}
It is apparent from Eq.\ref{eq:tempf} which is derived combining Eq.\ref{eq:dcap} and Eq.\ref{eq:drad}, that the late-time heating is decided by the DM capture rate, in turn by the DM-neutron scattering cross-section for a given local DM density. Its dependence on the DM mass is mild via $\sigma_g$ only, apparent from Eq.\ref{eq:cap}.
 
Returning to the condition of forming a BEC state, it can also be recast in terms of the ground state abundance of DM given by,
\begin{align}
N^0_{\chi}(t)= N_\chi(t)\bigg[1-\left(\frac{T_{NS}}{T_c}\right)^{3/2}\bigg].
\end{align} 

To note, for sufficiently large acquisition of DM particles inside the NS,  $T_c$ can be larger than $T_{NS}$ even for young NSs. However, it should be ensured that for such a large accumulation, DM particles do not become self-gravitating. The self-gravity of DM particles is effective when the DM density, $\rho_\chi \gtrsim \rho_c$ in a region with radius $r_{th}$. Consequently, the dark core collapse happens, rather than forming a BEC. The number of DM particles needed for the onset of the self-gravitation is given by \cite{Acevedo:2020gro}
\begin{align}
N_{sg}(t)=\frac{4}{3}\pi r^3_{th}\frac{\rho_c}{m_\chi}.
\label{eq:nsg}
\end{align}  
The BEC state forms only when $T_{NS}<T_c$ and $N_\chi(t)< N_{sg}(t)$ are simultaneously satisfied. The self-gravity and the BEC formation brings two timescales or temperature scales for thermalized DM particles, namely $T_c$ and $T_{sg}$. The characteristic temperature scale for the onset of self-gravitation, $T_{sg}$ is given by
\begin{align}
T_{sg}= 9\times 10^{-8}~ {\rm MeV} \left(\frac{N_\chi}{10^{36}}\right)^{2/3} \left(\frac{m_\chi}{10^3~ {\rm GeV}}\right)^{5/3}\left(\frac{5\times 10^{38}~ {\rm GeV~ cm^{-3}}}{\rho_c}\right)^{2/3}.
\end{align}  
Noticeably, $N_{sg} (t)$, $T_{sg}(t)$ are too functions of time, which make our dynamical analysis essential. In terms of characteristic temperatures BEC forms when $T_{sg}<T_{NS}<T_c$. DM particles become self-gravitating without forming a BEC when $T_c<T_{NS}<T_{sg}$. 

\subsection{Black hole formation inside a neutron star}
BH formation driven by the dark core collapse inside a NS is heavily dependent on the thermal states of DM particles. For a BEC state, as most of particles are in the ground state, i.e. corresponding to the vanishing momentum state, the collapse to a BH can be achieved with less number of particles compared to a non-BEC state that involves substantial contribution from non-vanishing momenta. In the absence of any self-interaction, DM particles in the ground state collapse under its own gravity when its number exceeds the \textit{Chandrasekhar} limit, i.e.  
\begin{align}
N_{ch}= \frac{16 M^2_P}{m^2_\chi}.
\end{align}
where $M_P =2.34\times 10^{18}$ GeV. For a BEC state, $N_\chi\simeq N^0_\chi$ and the BH forms when $N^0_\chi > N_{ch}$.  For a non-BEC state, DM particles becomes self-gravitating and thereafter leads to the BH formation when \cite{Acevedo:2020gro}
\begin{align}
N_\chi > Max[N_{sg},N_{ch}].
\end{align}
In the present analysis, $N_{sg}$ being always greater than $N_{ch}$, decides the BH formation. The BH formation from a non-BEC state, as wee see, is more demanding due to pressure support from the non-zero momentum states \cite{Bell:2013xk,Kouvaris:2012dz}. Consequently, the initial mass of a nascent BH differs depending on its thermal state. In particular, BH forming out of a BEC state has an initial mass, $M^i_{BH}= m_\chi N_{ch}$ whereas for a non-BEC state, $M^i_{BH}= m_\chi N_{sg}$. 

The continuous growth of the newly formed BH,leading to eventual destruction of the host star is decided by an interplay between the accretion of nuclear matter and evaporation of BH via Hawking radiation, governed by the equation \cite{Kouvaris:2011fi}
\begin{align}
\frac{dM_{BH}}{dt}= \frac{4\pi \rho_c G^2 M^2_{BH}}{c^3_s}-\frac{1}{15360\pi G^2 M^2_{BH}}.
\end{align}
Here $c_s (=0.17 c)$ is the speed of sound inside the star. The nascent BH starts growing if $dM_{BH}/dt >0$, i.e. the accretion is larger than the evaporation rate, which in turn gives the following lower bound on the initial mass of BH with negligible Hawking radiation.
\begin{align}
M^i_{BH} \gtrsim 5.7\times 10^{36} ~{\rm GeV}.
\label{eq:bheva}
\end{align}
With this one can estimate the \textit{collapse time} \cite{Singh:2022wvw}, within which the small black hole acquires   more baryonic matter and eventually devour the entire host star, forming a BH with a mass, equal to the NS mass. The collapse time is calculated from Eq.\ref{eq:bheva} ignoring the second term as,
\begin{align}
t_c \simeq \frac{c^3_s}{4\pi \rho_c G^2 M^i_{BH}} \simeq 40 ~ {\rm years} \left(\frac{10^{42} ~{\rm GeV}}{M^i_{BH}}\right).
\label{eq:collapse}
\end{align} 
As is evident that $t_c$ depends on the initial mass of the BH, which is determined by primeval thermal state of DM particles, leading to the BH formation. Therefore, the collapse time for BEC and non-BEC cases are quite different, as apparent in the following.
\begin{align}
t_c = 
\begin{cases}
4.6 \times 10^5 ~{\rm years} ~\left(\frac{m_\chi}{\rm GeV}\right) ~~ \text{for BEC BH},\\
3.6\times 10^{2} ~{\rm years} ~\left(\frac{m_\chi}{10^4~\rm GeV}\right)^{3/2} \left(\frac{10^{-5}~\rm MeV}{T_{NS}}\right)^{3/2} ~~ \text{for non-BEC BH}.
\end{cases}
\label{eq:colltime}
\end{align}
This difference becomes instrumental in distinguishing a BEC state BH from the other in the gravitational wave signal related to the binary system. We have briefly discussed the distinction between these two kinds BHs originated from collapse of NSs in Sec.\ref{sec:sec4}.
\section{Results and Discussions}
\label{sec:sec3}
In this section, we simultaneously solve the evolution equations of NS temperature (Eq.\ref{eq:temp}) and the yield of DM particles (Eq.\ref{eq:number}) inside the NS residing in a DM halo with $\rho_\chi =10^3 ~{\rm GeV~ cm^{-3}}$, for given DM parameters : DM-neutron scattering cross-section ($\sigma_{\chi n}$), DM mass ($m_\chi$) and DM annihilation cross-section ($\expval{\sigma v}$). We work with initial conditions as, $T^i_{NS}=1$ MeV and $N^i_\chi=0$. The NS temperature becomes only a function of time within first $100$ years of its birth, therefore we start our analysis from $t=100$ years with $T_{NS}=1$ MeV, consistent with standard cooling mechanism of a NS. Subsequently, we determine the thermal state of DM as well as the stability of the NS. Our goal is to find the DM parameter space allowed by observations of old NSs. As we will see, for some part of the parameter space DM particles will form either a BEC or a non-BEC state. For the subset of this parameter space, the host star transmutes to a BH. We present our results separately for annihilating and non-annihilating DM scenarios. All the results are shown for typical NS parameters, namely $M_{NS}=1.4 M_\odot$ and $R_{NS}=10.6$ km. Given the value of $R_{NS}$, the NS mass can vary within $(0.8-2.2) M_\odot$ \cite{Bramante:2023djs}, which modifies the DM capture rate by less than factor of 2 as compared to our present choice of NS parameters.
\begin{figure}[htb!]
\hspace{-1.8cm}
\includegraphics[scale=0.45]{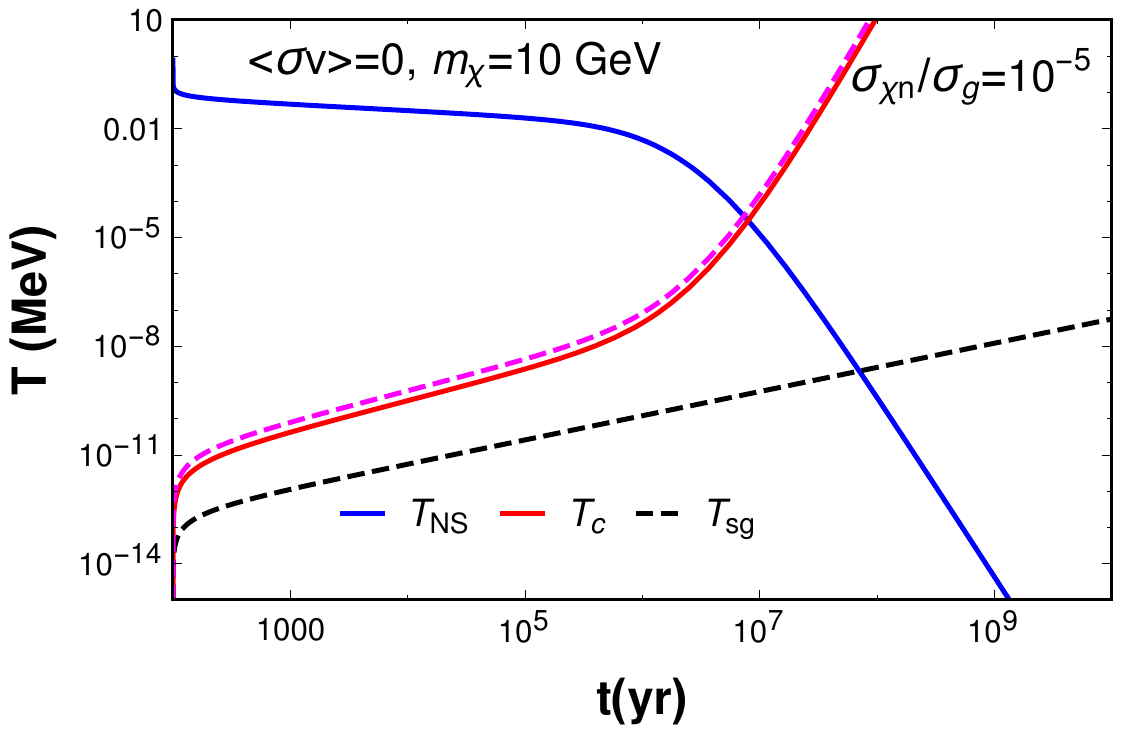}
\includegraphics[scale=0.45]{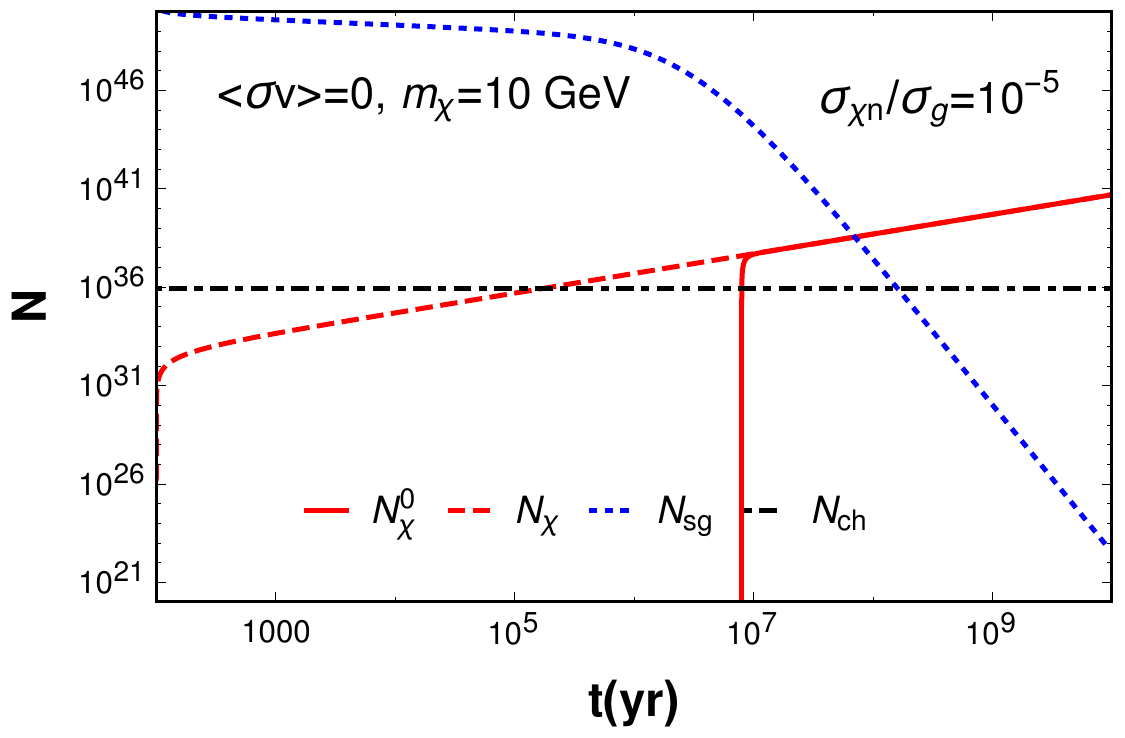}
\caption{Evolution of different temperature variables (left panel) and DM number variables (right panel) for non-annihilating DM scenario, for which the BEC forms. The present choice of parameters, shown by ($\bm{\star}$) symbol in Fig.\ref{fig:scan} is excluded from the observation of the old NS.}
\label{fig:tc}
\end{figure}

\subsection{Case I : Non-annihilating dark matter}
To illustrate the effects of accumulation of DM particles, we first choose $\sigma_{\chi n}/\sigma_g =10^{-5}$, $m_\chi=10$ GeV and $\expval{\sigma v}=0$, for which DM particles settle into a BEC state. In the left panel of Fig.\ref{fig:tc}, we have shown the evolution of NS temperature ($T_{NS}$) (blue solid line), the critical temperature ($T_c$) (red solid line) and the characteristic temperature for self-gravitation ($T_{sg}$) (black dashed line). In the absence of DM annihilation, $T_{NS}$ follows the standard cooling procedure, which initiates rapid cooling after $t=10^6$ years. On the other hand, $T_c$ increases gradually due to continuous acquisition of DM particles. Thus, $T_{NS}$ becomes smaller than $T_c$ around $t=8\times 10^6$ years,  which  marks the BEC formation. We note that the effect of self-gravitation remains small until the BEC formation, indicated by $T_{sg}$ being smaller than $T_c$ at the relevant epoch. Consequently, the BH formation is to be determined by the number of DM particles, exceeding the Chandrasekhar limit ($N_{ch}$). Thus the current choice of parameters, indicated by ($\bm{\star}$) mark in Fig.\ref{fig:scan} is excluded from the observation of an old NS. As an aside, we note the evolution of $T_c$ is mostly unaffected with a different choice of the NS mass, i.e. $M_{NS}= 2.2 M_\odot$, keeping the radius and baryon density fixed as mentioned earlier. This has been shown by the magenta dashed line in the left panel of Fig.\ref{fig:tc}.

In the right panel of Fig.\ref{fig:tc}, the BEC formation is reasserted with the ground state population ($N^0_\chi$) (red solid line) shooting up to a value, equal to the total DM abundance ($N_\chi$) (red dashed line) for $T_{NS}\leq T_c$. For $T_{NS}>T_c$, most of the particles are in non-zero momentum states, which contribute to substantial pressure before the BEC formation. Consequently, BH does not form albeit $N_\chi$ continues to grow since the initial epoch and surpasses $N_{ch}$ at $t\approx 10^5$ years. The self-gravity of DM particles can not destabilize the dark core as long as $N_\chi<N_{sg}$. In the present case, the collapse happens only when the BEC forms, for which $N_{sg}$ becomes an irrelevant quantity for BH formation.

The above-mentioned analysis is self-consistent only if DM particles are thermalized within the relevant epoch. We estimate the thermalization time ($t_{th}$) at the BEC formation epoch using Eq.\ref{therm}. For $\sigma_{\chi n}/\sigma_g =10^{-5}$ and $m_\chi=10$ GeV, we get $t_{th}=0.08$ years, which ensures DM particles become thermalized well within the formation of BEC. Suffice to say, the thermalization time-scale depends on the ambient NS temperature, which is determined dynamically in our analysis.
\begin{figure}[H]
\hspace{-1.8cm}
\includegraphics[scale=0.46]{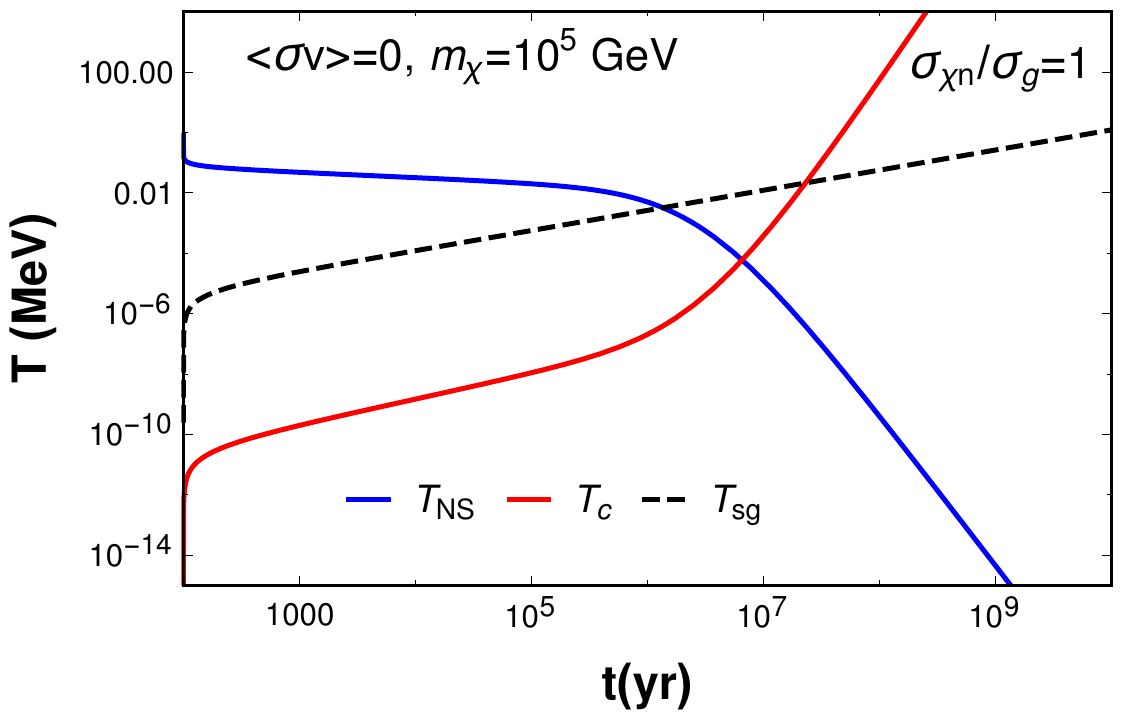}
\includegraphics[scale=0.45]{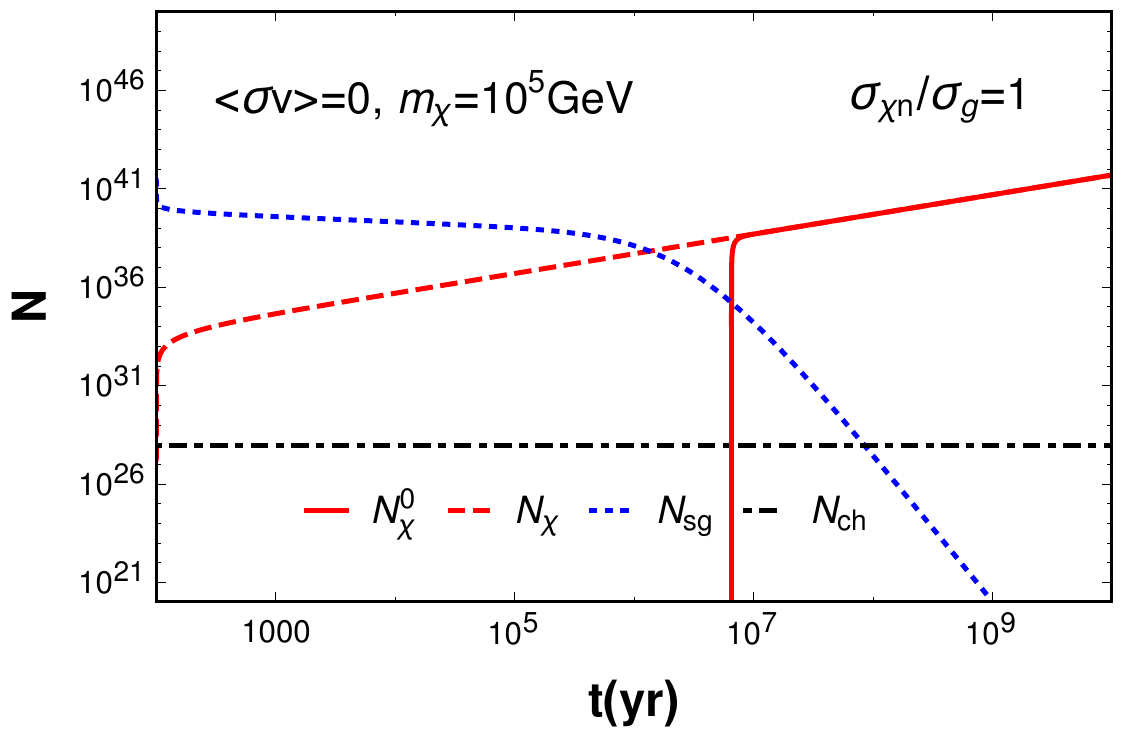}
\caption{Same as in Fig.\ref{fig:tc}, but with different choice of parameters for which DM particles become self-gravitating rather than forming a BEC. The present choice of parameters, shown by ($\bm{\times}$) symbol in Fig.\ref{fig:scan} is excluded from the observation of the old NS.}
\label{fig:tc1}
\end{figure}

In Fig.\ref{fig:tc1} we show that DM particles become self-gravitating rather than forming a BEC for parameters, $m_\chi=10^5 ~{\rm GeV}$ and $\sigma_{\chi n}/\sigma_g=1$, which have also been shown in Fig.\ref{fig:scan} by ($\bm{\times}$). In the left panel, $T_{NS}$ falls below $T_{sg}$ around $t=6\times 10^6$ years, at which $T_c$ remains smaller than both $T_{NS}$ and $T_{sg}$. This signifies the imminent collapse of DM particles to a BH without being in a BEC state. It is also evident from the right panel plots that $N^0_\chi$ remains vanishing when $N_\chi$ takes over $N_{sg}$. In this scenario, $N_{sg}$ is the deciding factor of BH formation, unlike the former case where $N_{ch}$ is the relevant quantity. To note, the BH formation via self-gravitation is favorable for heavy DM scenarios, as evident from Fig.\ref{fig:scan}.
\begin{figure}[htb!]
\centering
\includegraphics[scale=0.5]{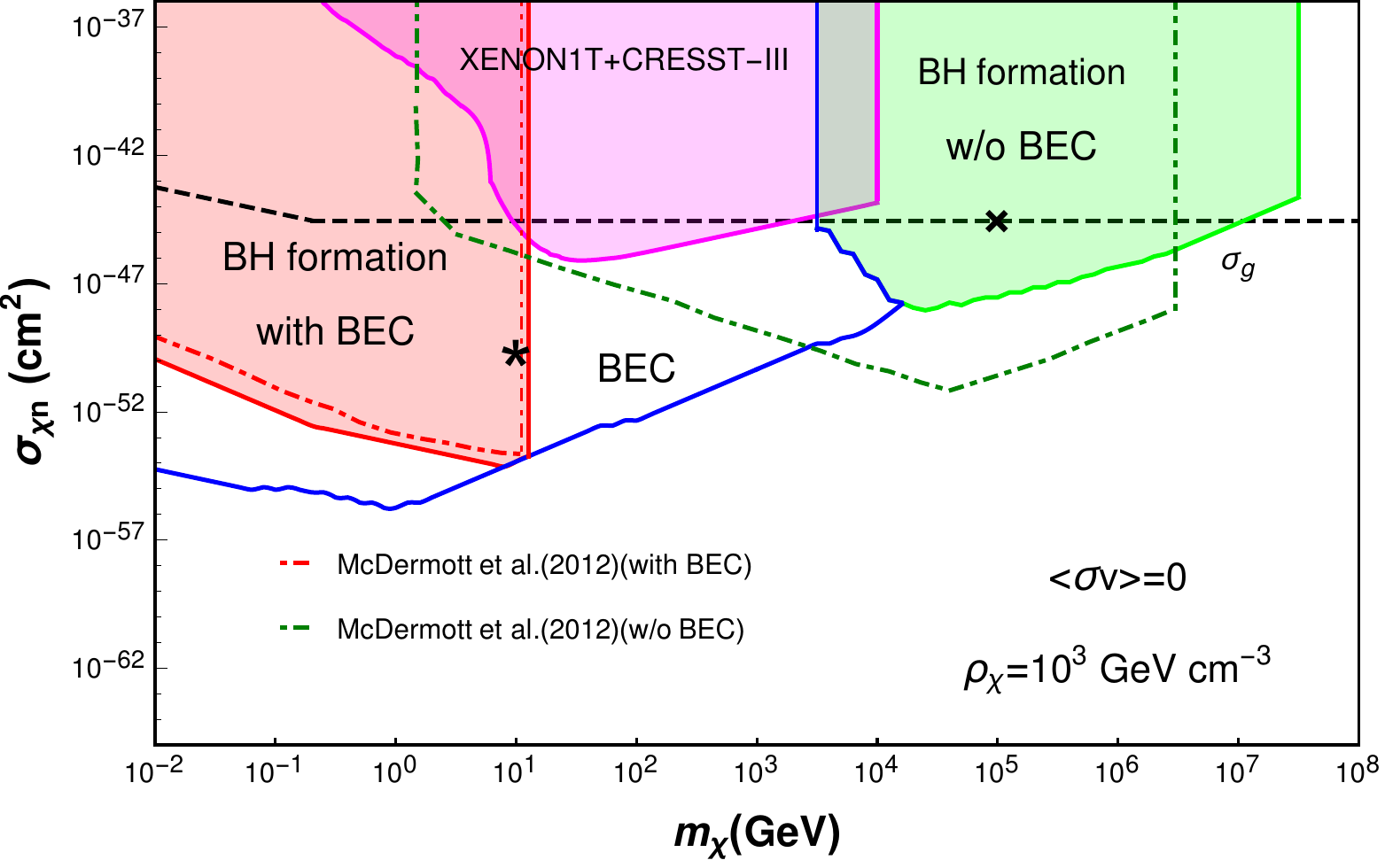}
\caption{\textit{Depiction of DM parameter space for non-annihilating DM scenarios}: The black dashed line denotes the geometric limit ($\sigma_g$) of the DM-neutron scattering cross-section ($\sigma_{\chi n}$).  The region surrounded by the blue solid line denotes relevant parameter points for BEC formation. The red and green shaded regions are excluded given the observation of an old neutron star with a typical age of $10^{10}$ years. For parameters shown by ($\bm{\star~,~\times}$) symbols, we have shown the evolution of different temperature variables and DM number variables in Fig.\ref{fig:tc} and Fig.\ref{fig:tc1}, confirming the distinction between a BEC and a non-BEC state. The magenta shaded region signifies the parameter space, already excluded by the current direct detection experiments. We contrast our results from the dynamical analysis with that (indicated by the red dot-dashed and the green dot-dashed lines) of Refs.\cite{McDermott:2011jp,Garani:2018kkd}, which considers a fixed NS temperature, $T_{NS}=10^{-5}$ MeV. For details see the main text.}
\label{fig:scan}
\end{figure}

Now, we study relevant DM parameter space in which distinction between a BEC and a non-BEC state, followed by BH formation becomes apparent. In Fig.\ref{fig:scan} we show the regions of BEC formation (left upper corner, surrounded by the blue solid line),  BH formation out of a BEC state (the red shaded region)and BH formation without a BEC state (the green shaded region) in the plane of DM mass and DM-neutron scattering cross-section, covering DM masses from $10~$MeV to $10^8$ GeV. The rest of the parameter space indicates the region where neither a BEC state forms nor a BH. For $m_\chi \lesssim 10$ MeV, the evaporation effects are important as apparent from Eq.\ref{eq:eva}, at the initial phases of the NS temperature evolution. We have neglected the effect throughout our dynamical evolution, thus comes the lower limit on the DM mass to be $m_\chi\gtrsim 10$ MeV. For DM masses $10~{\rm MeV}\leq m_\chi \leq 16 ~{\rm TeV}$, the BEC state forms out of captured DM particles, which is depicted by the region surrounded by the blue solid line as mentioned earlier. In the rest of DM parameter space shown in Fig.\ref{fig:scan}, BEC does not form. In red and green-shaded regions, the BH forms at the center of the NS from the accumulated non-annihilating DM. Subsequently, the BH continues to grow and devours the whole NS, thereby comes the relevant constraint, details of which are the following.

In globular cluster like $M4$, PSR B1620-26, a millisecond pulsar has been observed with an age of $10^{10}$ years \cite{Camilo:2005aa,Kouvaris:2010jy}. In this environment, we can have DM-dense regions with $\rho_\chi =10^3 ~{\rm GeV~ cm^{-3}}$ \cite{Bertone:2007ae,McCullough:2010ai}. Hence, observation of an old NS indicates that the NS has not been converted into a BH despite the accumulation of DM particles over its lifetime. It then allows us to exclude the relevant DM parameter space shown by the red and green shaded regions in Fig.\ref{fig:scan}. In particular, older is the neutron star and denser is the DM environment - stronger becomes the constraint on $\sigma_{\chi n}$, which motivates our present choice of the NS  and the DM environment. 

In the red shaded region BH forms and continues to grow, subject to the erstwhile BEC state. In this case, for $m_\chi \geq 12$ GeV, $\sigma_{\chi n}$ becomes unconstrained. This is because of newly formed BH with an initial mass, $M_{BH,i}\sim m_\chi N_{ch}$ evaporates according to Eq.\ref{eq:bheva}. In the region with $m_\chi \lesssim 12$ GeV, the constraint on $\sigma_{\chi n}$ gets stronger as the DM mass increases in general. However, there is some subtlety about the upper limit on $\sigma_{\chi n}$, varying over the relevant DM mass range. In particular, for $m_\chi <0.2$ GeV, the upper limit is decided as $\sigma_{\chi n}\propto m^{-2}_\chi$, whereas for $m_\chi >0.2$ GeV, $\sigma_{\chi n}\propto m^{-1}_\chi$. This can be understood from the following equation after solving Eq.\ref{eq:cap}, Eq.\ref{eq:geom} and Eq.\ref{eq:number}, demanding $N_\chi=N_{ch}$ for the gravitational collapse. 	
\begin{align}
 \sigma_{\chi n}m_\chi  {\rm Min}\left[\frac{m_\chi}{0.2 ~{\rm GeV}},1\right] =\text{const}.
\end{align}

The green shaded region shows the self-gravitating phase leading to the formation of BH. In this case, the constraint on $\sigma_{\chi n}$ becomes weaker due to larger number of DM particles requiring to destabilize a NS. To contrast with the former case, the upper limit on $\sigma_{\chi n}$ increases as the DM mass increases in the regime, $m_\chi \gtrsim 10$ TeV. In this region, BH formation is governed by $N_{sg}$, which depends on both $T_{NS}$ and $m_\chi$ (see Eq.\ref{eq:nsg}), unlike the formation of BH in the BEC case, which depends only on $m_\chi$. Consequently, the initial mass of the nascent BH is given by, $M_{BH,i} \sim m_\chi N_{sg}\propto (T_{NS}/m_{\chi})^{3/2}$, which again necessitates our dynamical analysis. Similar to the BEC case, the BH evaporation effects open up the DM parameter space but for $m_\chi \gtrsim 3\times 10^7$ GeV. 

In passing, we also note that in the high mass end, the upper bound on $\sigma_{\chi n}$ exceeds its geometric limit, $\sigma_g$ shown by the black dashed line in Fig.\ref{fig:scan}. Although beyond the geometric limit the capture rate becomes fixed at its highest value, higher rates of DM-neutron interaction is necessary for DM thermalization, which validates our calculation for BH formation out of a thermal state.

In addition to NS constraints, we present the current direct detection bound from XENON1T\cite{XENON:2019gfn} (for GeV to TeV DM) and CRESST-III \cite{CRESST:2019jnq} (mainly for sub-GeV DM) on the $\sigma_{\chi n}-m_\chi$ plane, which is somewhat complementary to the current exclusion limits coming from DM capture scenarios. It is worth noting that  $m_\chi \in [12 {~\rm GeV},3 {~\rm TeV}]$, DD experiments remain instrumental in constraining the DM-neutron cross-section. This is an important implication of non-annihilating bosonic DM forming a BEC state, thereby evading constraints from the NS observations often advocated to probe smaller DM-neutron cross-sections in the literature. This result is quite robust with varying the local DM density, as the concerned mass range is dependent on the NS properties. In particular, the lower limit, i.e. $m_\chi = 12$ GeV, beyond which all associated nascent BHs evaporate is fixed given the baryon density and the sound velocity inside the NS. The upper limit, i.e. $m_\chi = 3$ TeV, beyond which BH forms due to self-gravity, is again dependent on the baryon density of the NS. However, for sub-GeV DM and DM with masses larger than tens of a TeV, NS observations can probe various DM scenarios with greater sensitivity compared to the DD experiments, albeit various uncertainties regarding NS parameters.

Now, let us compare the present result in the context of already existing bounds in most of the literature \cite{McDermott:2011jp,Bramante:2013hn,Garani:2018kkd,Bhattacharya:2023stq,Lu:2024kiz}. In these papers, the constraints on the $\sigma_{\chi n}-m_\chi$ plane are derived considering BEC and non-BEC state separately with a fixed NS temperature. In our work, we dynamically determine the demarcation between a BEC and a non-BEC state given a set of DM parameters. This is one of the main differences from the earlier works. Additionally, in our analysis we check thermalization of DM particles throughout the evolution, whereas in previous studies the thermalization timescale is derived considering a fixed final temperature of a NS. However, the upper bound on $\sigma_{\chi n}$ coming from the BH formation in the BEC case, does not disagree to a great extent comapred to these works. The results for the BEC and non-BEC states from Fig.1 of Ref.\cite{McDermott:2011jp} have been shown by the red dot-dashed and green dot-dashed line respectively in our Fig.\ref{fig:scan}. It is apparent that the significant difference emerges for the non-BEC case. We have checked that the recent analysis in Ref.\cite{Garani:2018kkd} (depicted in Fig.10) agrees with results shown in Ref.\cite{McDermott:2011jp}, thereby our results differ from both of the references. 

In particular, there are three differences for the BH formation from a non-BEC state. Firstly, the upper bound on $\sigma_{\chi n}$ in the mass regime, $1 ~{\rm GeV} \lesssim m_\chi \lesssim 5 ~{\rm TeV}$ is not applicable due to BEC formation, as implied by our analysis. Moreover, this region is not even ruled out (claimed otherwise in Ref.\cite{McDermott:2011jp, Garani:2018kkd}) due to BH evaporation. Secondly, the constraint on $\sigma_{\chi n}$ as derived from our dynamical treatment is appreciably different for $m_\chi > 16 ~{\rm TeV}$. This stems from the fact that the DM number needed for the onset of self-gravitation is significantly higher in our analysis than in the previous studies. In fact, given the DM parameters, the temperature, $T_{NS}$ at which the self-gravitation sets in, is completely determined. For example we see for $m_\chi=10^5$ GeV, $N_{sg}=N_{\chi}$ at $T_{NS}\approx 0.005 ~{\rm MeV}$, apparent from Fig.\ref{fig:tc1}. In previous studies, the onset of self-gravitation is assumed at $T_{NS}=10^{-5}~{\rm MeV}$, which leads to the under-estimation of DM number required for the collapse. Therefore, in our analysis, the self-gravitating regime in the non-BEC case sets in earlier, which requires larger $\sigma_{\chi n}$. Finally, the nascent BH mass becomes greater in our analysis for the same DM mass, which has been reflected in the higher cut-off in the DM mass related to BH evaporation. In particular, the DM mass cut-off, beyond which there is no constraint from the stability of a NS is found to be $10$ times larger than the non-dynamic analyses.

\subsection{Case II : Annihilating dark matter}

In this section, we shall study the effects of DM annihilation in the formation of a BEC state and subsequently collapse to a BH. In presence of DM annihilation, the higher DM capture rate is expected for BEC and BH formation, compared to the non-annihilating case. In addition, if the annihilation products, such as charged leptons and quarks, are trapped inside the NS, the NS temperature departs from its standard evolution. Due to energy injection to the NS, $T_{NS}$ becomes constant at late epochs ($t\gtrsim 10^7$ years), leading to the late-time heating as mentinoed in Sec.\ref{sec:sec2}. The annihilation products get trapped due to dominant electromagnetic interaction with electrons present in the NS crust. The SM neutrinos with energy, $E_\nu \gtrsim \mathcal{O}(\rm{MeV})$, produced from DM annihilation can also be trapped due to weak interaction with electrons \cite{Shapiro:1983du,Cermeno:2017ejm} and contribute to the late-time heating. If DM annihilates to sterile neutrinos or some long-lived mediators with negligible coupling to SM particles, then these particles escape the NS without any subsequent heating effects \cite{Leane:2021ihh,Bose:2021yhz,Nguyen:2022zwb}. In the present study, we consider the DM annihilation only to SM states, leading to the  heating effect for DM masses, $m_\chi \geq 10$ MeV, as in the previous section. Moreover, we restrict our discussion to the s-wave contribution of the thermally averaged cross-section ($\expval{\sigma v}$) of DM annihilations. 

\begin{figure}[htb!]
\hspace{-1.8cm}
\includegraphics[scale=0.45]{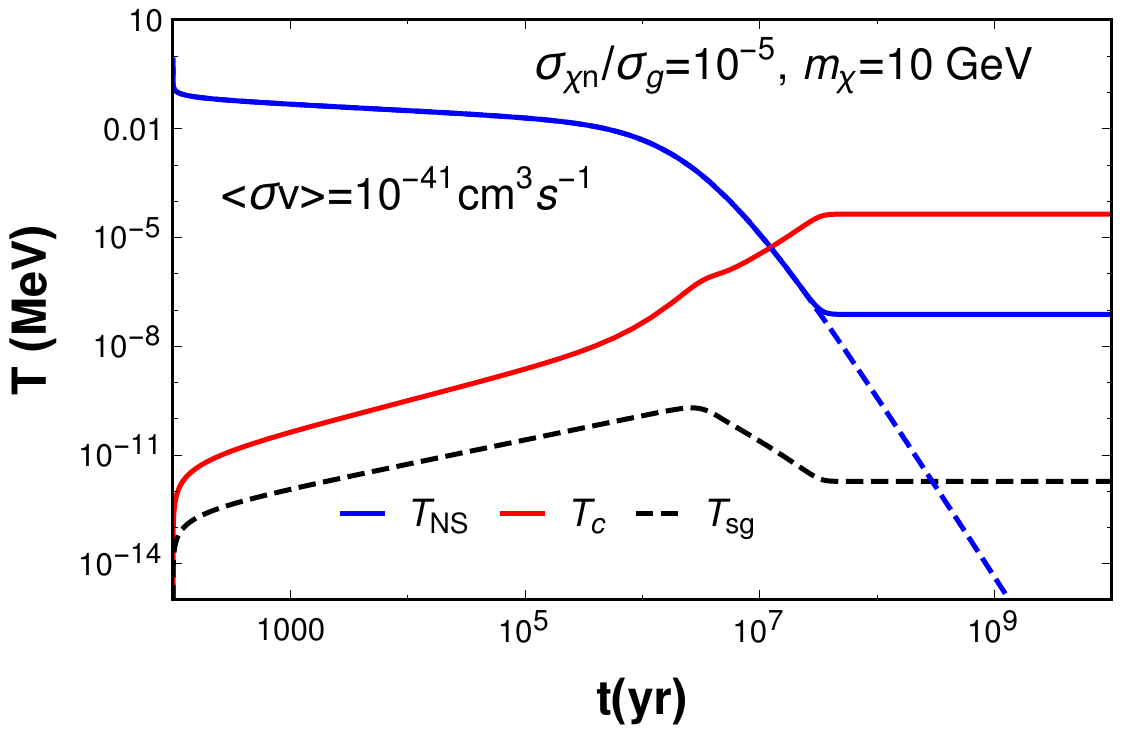}
\includegraphics[scale=0.45]{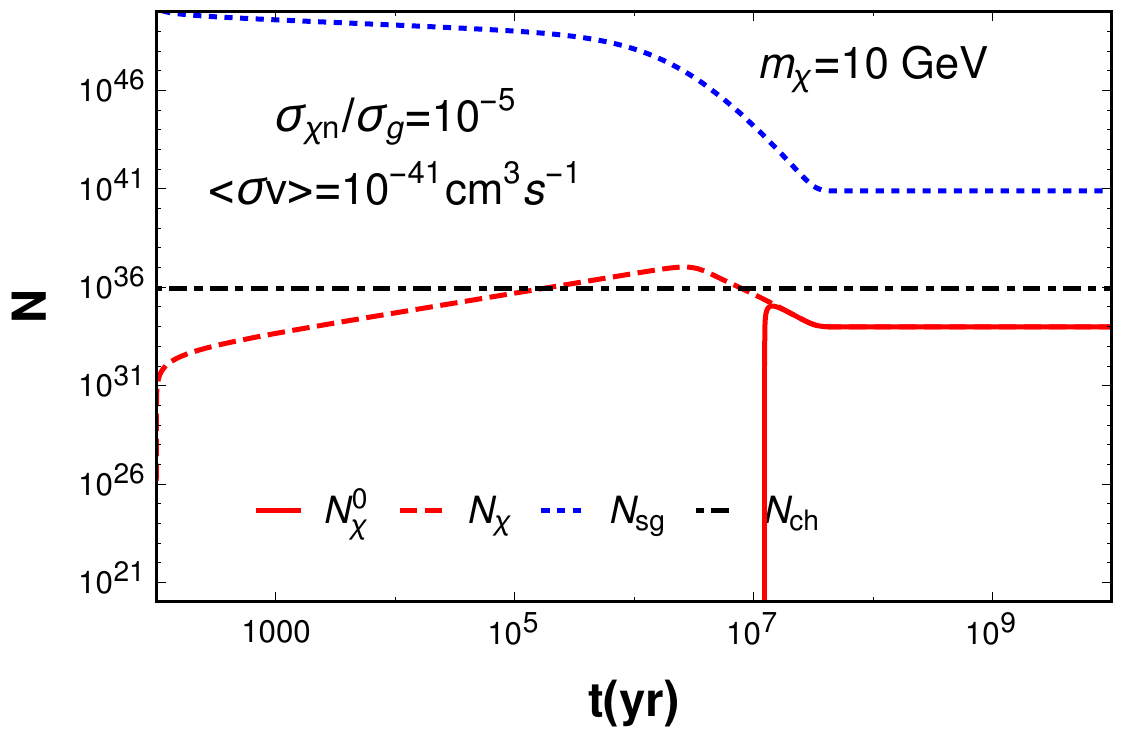}
\caption{Same as in Fig.\ref{fig:tc}, but for annihilating DM scenario, with $\expval{\sigma v}=10^{-41} {\rm cm^3 ~s^{-1}}$. The rest of DM parameters are set as in Fig.\ref{fig:tc1}, leading to the BEC formation as before, while BH does not form due to sufficient DM annihilation.}
\label{fig:atc}
\end{figure}

In Fig.\ref{fig:atc} we demonstrate the annihilation effects for the BEC case with $\expval{\sigma v}=10^{-41} {~\rm cm^3 ~s^{-1}}$. To contrast with the non-annihilating case, we set all other DM parameters at values as in Fig.\ref{fig:tc}, i.e. $m_\chi=10$ GeV, $\sigma_{\chi n}/\sigma_g=10^{-5}$. In the left panel, $T_{NS}$ denoted again with blue solid line \textit{freezes out} with a value, $T^f_{NS}=7.6\times 10^{-7}$ MeV after $t=3\times 10^7$ years. To note, in the non-annihilating DM scenarios $T_{NS}$ (shown by blue dashed line) continues to decrease over time. We have checked that the freeze-out value, $T^f_{NS}$ is well approximated by Eq.\ref {eq:tempf}, as the capture-annihilation equilibrium is ensured with the current choice of annihilation cross-section. With the evolution of $N_\chi$ (red dashed line) shown in the right panel of Fig.\ref{fig:atc}, the capture-annihilation equilibrium becomes apparent. $N_\chi$ increases due to dominant capture events till $t\simeq 10^6$ years, after which the annihilation term in Eq.\ref{eq:number} dominates. In particular, the annihilation rate is a temperature dependent quantity (see expression of $C_a$ in Sec.\ref{sec:sec2}) and continues to grow as long as $T_{NS}$ decreases. Hence, $N_\chi$ depletes for a significant amount of time, i.e. $10^6\lesssim t \lesssim 3\times 10^7~{\rm years}$. The annihilation rate soon becomes constant, as $T_{NS}$ reaches its freeze-out value. Subsequently, the capture-annihilation equilibrium is achieved, leading to the freeze-out of the DM abundance as well. The freeze-out value of $N_\chi$ is determined by  the DM capture and DM annihilation rate as, $N^f_\chi=\sqrt{C_c/C_a}$.  

Now due to annihilation, the BEC formation in this case happens (marked by $T_{NS}< T_c$) around $t=1.2\times 10^7$ years, which is a bit later than in the non-annihilation case, i.e. $t= 8\times 10^6$ years. However, the BEC formation does not lead to the subsequent BH formation as $N^0_\chi$ remains smaller than $N_{ch}$, unlike its non-annihilating counterpart - see the right panel of Fig.\ref{fig:atc}. The number required for self-gravitation remains larger than the ambient DM abundance throughout the evolution history, ensuring the consistency of the BEC formation as before.
\begin{figure}[htb!]
\hspace{-1.8cm}
\includegraphics[scale=0.7]{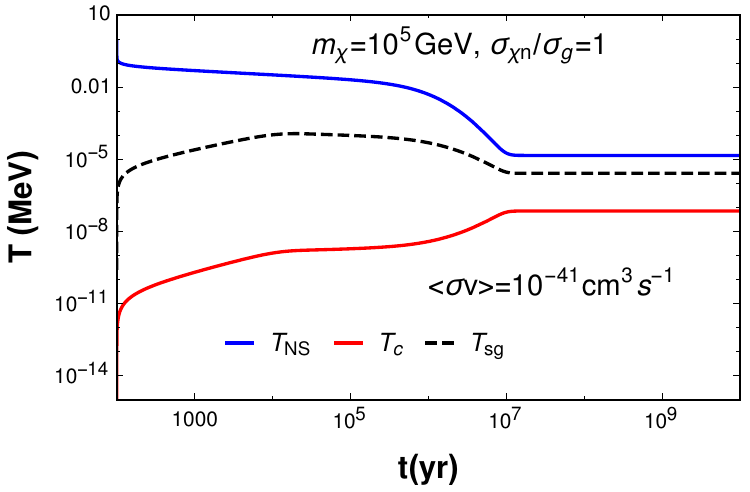}
\includegraphics[scale=0.45]{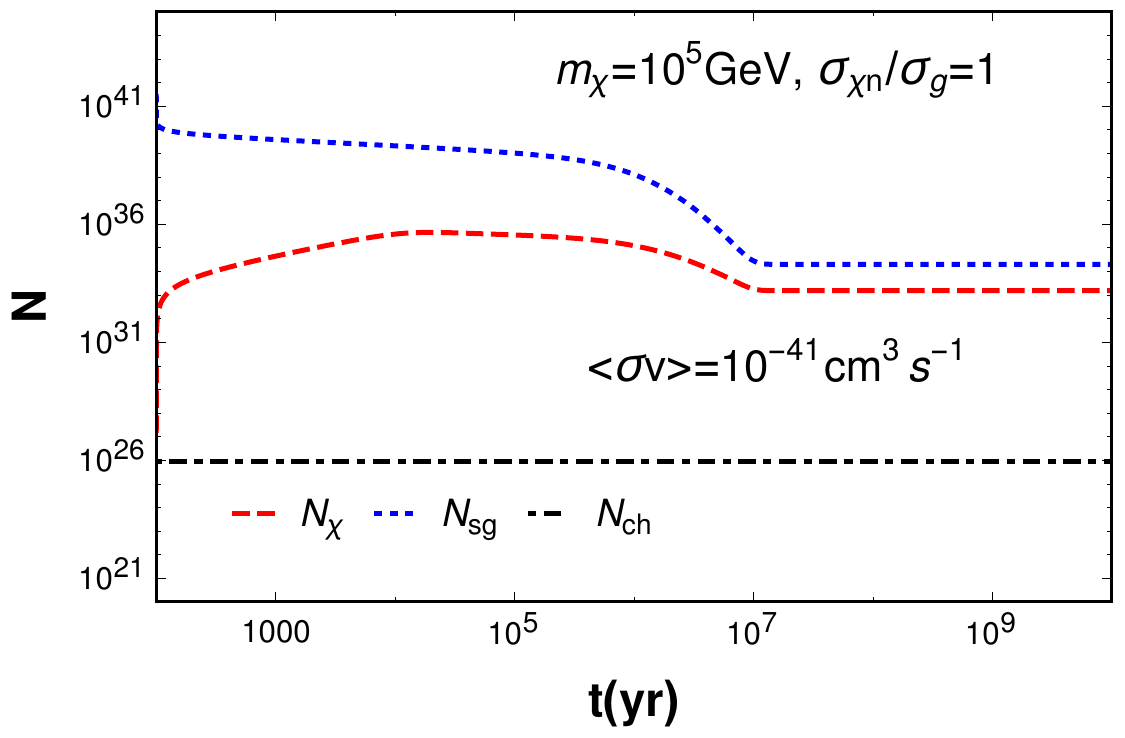}
\caption{Same as in Fig.\ref{fig:tc1}, but for annihilating DM scenario, with $\expval{\sigma v}=10^{-41} {\rm cm^3 ~s^{-1}}$. The rest of DM parameters are set as in Fig.\ref{fig:tc1}. Neither a BEC nor a BH form due to sufficient DM annihilation. }
\label{fig:atc1}
\end{figure}

It is evident from Fig.\ref{fig:atc1} that DM particles neither form a BEC nor become self-gravitating for the same parameters as in Fig.\ref{fig:tc1}. With $\sigma_{\chi n}/\sigma_g =1$, $T_{NS}$ is always larger than both $T_c$ and $T_{sg}$ , contrary to the case with $\expval{\sigma v}=0$.   In particular, for annihilating DM scenario, the higher DM capture rate leads to higher temperature for old enough NS, therefore both BEC formation and BH formation are hindered eventually.

\begin{figure}[htb!] 
\hspace{-1.8cm}
\includegraphics[scale=0.35]{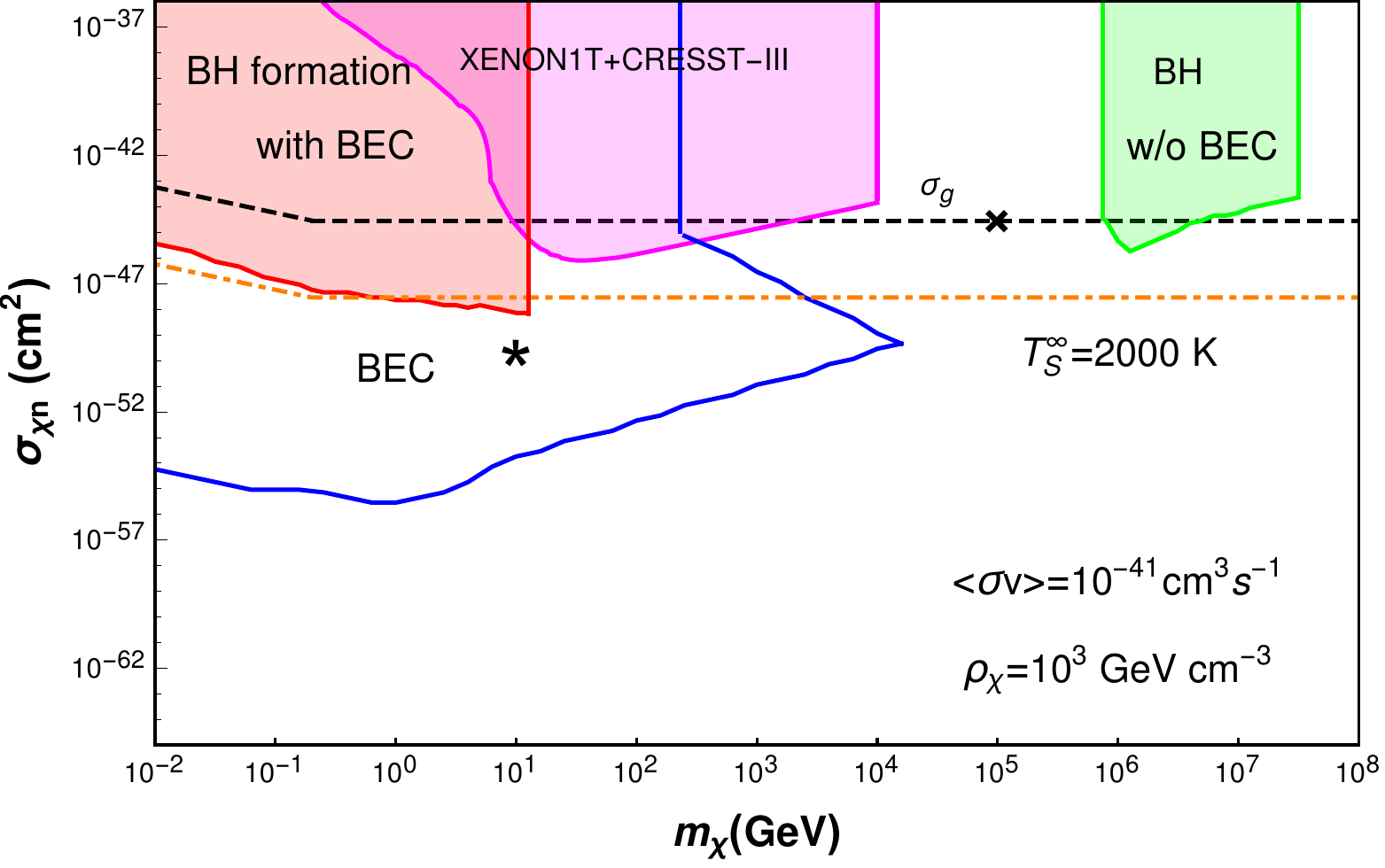}
\includegraphics[scale=0.35]{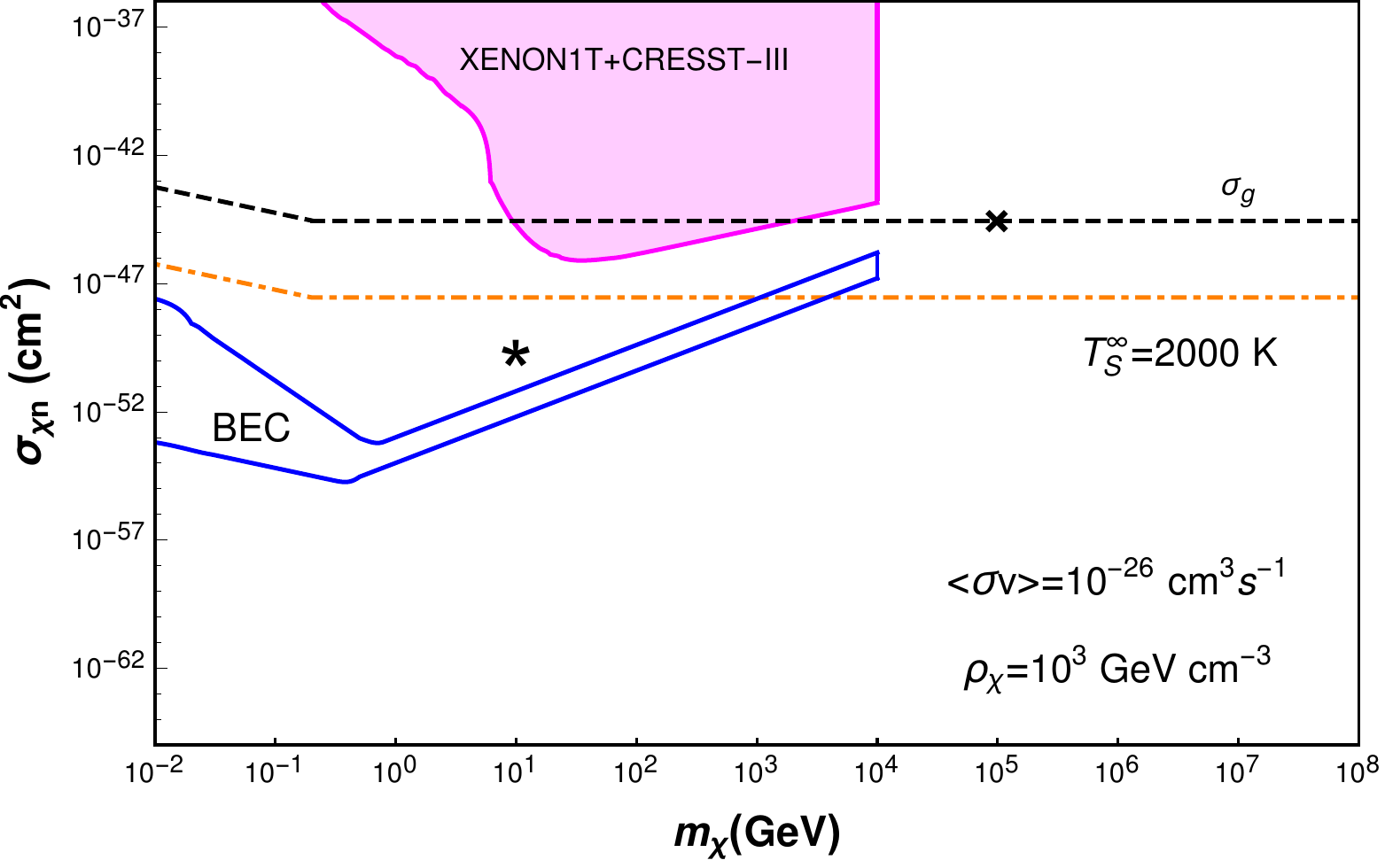}
\caption{\textit{Depiction of DM parameter space for annihilating DM scenarios}: Same as Fig.\ref{fig:scan}, but for two benchmark  values of DM annihilation rate : $\expval{\sigma v}=10^{-41}{~\rm cm^3 ~s^{-1}}$ (left panel) and $\expval{\sigma v}=10^{-26}{~\rm cm^3 ~s^{-1}}$ (right panel). For parameters shown by ($\bm{\star ~,~\times}$) symbols, BH does not form unlike the non-annihilating case in Fig.\ref{fig:scan}. The region above the orange dot-dashed line can be probed using the JWST observations of an old NS with $T^{\infty}_{S} \geq 2000$ K.}
\label{fig:scan1}
\end{figure}

As indicated in the previous two cases, the constraint on the plane of $\sigma_{\chi n}-m_\chi$ is expected to be modified significantly due to DM annihilation compared to non-annihilating case. Therefore, in Fig.\ref{fig:scan1} we show the relevant constraints for two benchmark values of DM annihilation rates, i.e. $\expval{\sigma v}=10^{-41}{~\rm cm^3 ~s^{-1}}$ and $\expval{\sigma v}=10^{-26}{~\rm cm^3 ~s^{-1}}$. The latter choice is motivated from the typical cross-section needed for saturating the relic density of a WIMP DM.  

The parameter region for BEC formation is greatly modified due to annihilation. This is clear by comparing the blue solid line surrounded regions in Fig.\ref{fig:scan} and the left panel of Fig.\ref{fig:scan1}.
BEC formation depends on both $\sigma_{\chi n}$ and $\expval{\sigma v}$. Increasing $\sigma_{\chi n}$ allows more accumulation of DM particles, thus expediting the BEC formation. On the other hand, due to annihilating nature of DM, more accumulation leads to heating up the NS. The interplay between these two effects creates the wedge-like structure within $500 {~\rm GeV} \lesssim m_\chi \lesssim 10 {~\rm TeV}$. In particular, when $\sigma_{\chi n}$ reaches above its geometric limit no further accumulation happens and it creates the vertical blue line in that region.
Now, for the case of annihilating DM  the BH forms within the BEC region for more restricted choices of $\sigma_{\chi n}$. In presence of annihilation smaller values of $\sigma_{\chi n}$  correspond to less number of accumulated DM particles, hindering subsequent BH formation. Note that, the maximum mass for which BH forms in the BEC state remains the same for both annihilating and non-annihilating DM. This is a consequence of BH evaporation only, which fixes the mass cut-off. Now, the BH formation from a non-BEC state takes place for $m_\chi > 5\times 10^5$ GeV for $\expval{\sigma v}=10^{-41} ~{\rm cm^3~ s^{-1}}$, significantly different from the former case depicted in Fig.\ref{fig:scan}. This is because the DM abundance is reduced via annihilation, thereby requiring heavier DM particles for the onset of self-gravitation. We emphasize that for parameter points shown in Fig.\ref{fig:scan} by ($\bm{\star ~,~\times}$) symbols, the BH does not form in the present case, contrary to the former one.

Now, we compare two panels of Fig.\ref{fig:scan1} to understand the effects of different values of DM annihilation cross-section. First of all, the parameter space of BEC formation gets shrunk considerably for $\expval{\sigma v}=10^{-26} {~\rm cm^3 ~s^{-1}}$. In particular, the minimum value of $\sigma_{\chi n}$ for BEC formation is higher than in the former scenario. This is because for the same value of $\sigma_{\chi n}$, more DM particles annihilate owing to larger $\expval{\sigma v}$. For instance, the parameter point shown by ($\bm{\star}$) in Fig.\ref{fig:scan1}, BEC state does not form in the right panel. Besides for DM masses ranging from $10$ MeV to $10$ TeV, there is a upper limit on $\sigma_{\chi n}$, above which BEC never forms, unlike the scenario shown in the left panel. This is resulted from the competing effects between the DM capture and annihilation, as explained earlier. The final distinction from the earlier scenario is that  BH never forms for such a large annihilation rate. Consequently, the constraint on $\sigma_{\chi n}-m_\chi$ plane is completely relaxed from the observation of old NSs.

However, DM annihilations allows us to probe $\sigma_{\chi n}$ for wide range of DM masses with the advent of modern telescopes. The thermal emissions from old NSs in the infra-red and visible frequencies can be detectable in telescopes, such as the JWST. In particular, JWST aims to detect NSs with the red-shifted surface temperatures ($T^{\infty}_S$) within 2000 K to 40000 K \cite{Raj:2024kjq,Chatterjee:2022dhp}, which can determine the cooling procedure of old NSs. Therefore, one can test various late-time heating scenarios related to astrophysics of NS, including the DM annihilation effects. Here, we use the sensitivity of the JWST to probe the DM-neutron cross-section, assuming the late-time heating coming only from DM annihilations. The typical surface temperature emerges in case of DM annihilation is around $10^{-6}$ MeV ($10^4 K$) for $T_{NS}=10^{-5}$ MeV, as apparent from Eq.\ref{eq:stemp}. In the both panels of  Fig.\ref{fig:scan1}, the region above the orange dot-dashed line can be probed with the detection of thermal emissions of 10 Gyr old NS having $T^{\infty}_S \geq 2000$ K. To remind, the surface temperature of the NS is related to $T^f_{NS}$, which is dependent on $m_\chi C_c$ (see Eq.\ref{eq:tempf}) in the capture-annihilation equilibrium. Consequently, $\sigma_{\chi n}$ is determined with mild dependence on the DM mass, $m_\chi$ through $\sigma_g$ apparent from Eq.\ref{eq:geom}. 

\begin{figure}[htb!]
\includegraphics[scale=0.5]{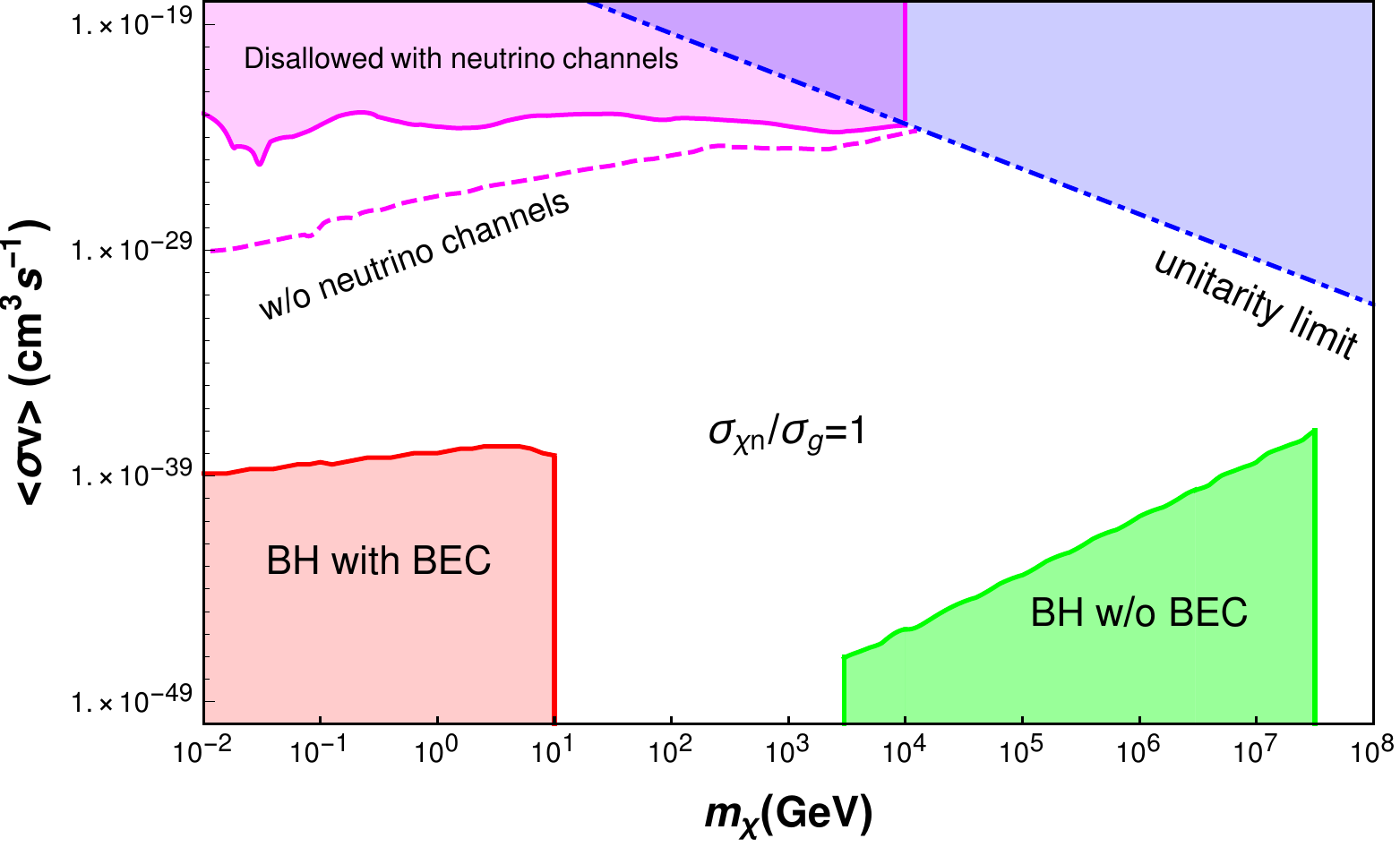}
\caption{Constraints on the s-wave contribution to the thermally averaged  cross-section ($\expval{\sigma v}$) of DM annihilations from indirect searches and the observation of old neutron stars. The constraint from the NS depends on the DM-neutron scattering cross-section, therefore we set $\sigma_{\chi n}/\sigma_g=1$ for illustration. We also show the unitarity bound on $\expval{\sigma v}$ for thermal DM scenarios.}
\label{fig:sigmav}
\end{figure}
 
In Fig.\ref{fig:sigmav} we compare the constraints on $\expval{\sigma v}$ stemming from the non-observation of any excess fluxes of gamma rays, electrons and neutrinos over standard astrophysical backgrounds in existing telescopes and neutrino detectors, with the constraints related to the observation of old NSs. The existence of an old NS sets a lower limit on $\expval{\sigma v}$, provided the DM capture rate is known. In particular, all values of $\expval{\sigma v}$ over its lower bound signify the substantial depletion of DM particles inside the NS, thereby not transforming the NS into a BH. Consequently, the red-shaded (for BEC state) and green-shaded (for non-BEC state) regions in Fig.\ref{fig:sigmav} are excluded for $\sigma_{\chi n}/\sigma_g =1$ and $\rho_\chi=10^3 {~\rm GeV ~cm^{-3}}$. The distinction between these two regions, again is coming from the BH formation, out of a BEC and a non-BEC state as shown in Fig.\ref{fig:scan}. For $m_{\chi} \lesssim 10 {~\rm GeV}$, in the BEC formation regime the lower bound on $\expval{\sigma v}$ is somewhat independent of DM mass, i.e. $\expval{\sigma v} \gtrsim 10^{-39}{~\rm cm^3~ s^{-1}}$. On the other hand, in the non-BEC regime the bound strengthens with increasing value of DM mass. This is due to the fact that  DM particles needed for BH formation (as shown in Eq.\ref{eq:nsg}) for a given capture rate decreases with the increase of $m_\chi$. Consequently, the BH still forms with larger DM annihilation rate, thereby strengthening the constraint on the $\expval{\sigma v}$. We note that for $\sigma_{\chi n}/\sigma_g < 1$, the bound on the $\expval{\sigma v}$ is expected to be weakened, as smaller capture rate demands smaller annihilation to form BH for both BEC and non-BEC DM states.

For $10~ {\rm GeV}\lesssim m_\chi \lesssim 5 {~\rm TeV}$ and $m_\chi\gtrsim 3\times 10^7$ GeV, there is no constraints on the $\expval{\sigma v}$ from the NS observations as nascent BHs evaporate via Hawking radiation, even in the non-annihilating DM scenario. We also note that the late-time heating of NS can not put any constraint on $\expval{\sigma v}-m_\chi$ plane as long as the capture-annihilation equilibrium condition is ensured. The minimum value of DM annihilation cross-section, necessary for the equilibrium is rather small, (e.g. $\expval{\sigma v}\gtrsim 10^{-58}{~\rm cm^3 ~s^{-1}}$ for $m_\chi=10^5$ GeV \cite{Bramante:2013hn}), compared to the present constraints shown in Fig.\ref{fig:sigmav}. 

In Fig.\ref{fig:sigmav} we also show the indirect detection bounds using results from Ref.\cite{Dutta:2022wdi}, including both - with neutrino channels (shown by the magenta solid line) and without neutrino channels (shown by the magenta dashed line) associated to the $s$-wave DM annihilation. In addition, we also show the unitarity limit (by blue dot-dashed line) on the thermal average of $2\rightarrow 2$ inelastic cross-section in the thermal freeze-out scenario for WIMPs, given by \cite{Griest:1989wd,Hui:2001wy,Bhatia:2020itt}
\begin{align}
\expval{\sigma v}_{uni}= \frac{4\sqrt{\pi}}{m^2_\chi}\sqrt{x_F},
\end{align}  
where $x_F =20$, the scaled time parameter in the DM freeze-out case. The unitarity limit indicates the maximum possible cross-section of DM annihilation for given value of $m_\chi$. It is apparent from Fig.\ref{fig:sigmav} that the region below the magenta lines is allowed and yet to be probed, with uncertainties originating from the different branching ratios of annihilation channels. It is apparent that one can reduce the parameter space of $\expval{\sigma v}$ using constraints from indirect searches and the NS observation simultaneously, once $\sigma_{\chi n}$ is known. This also shows a complementarity between direct and indirect searches of DM. 

\section{Outlook}
\label{sec:sec4}
Our present study is about bosonic DM particles forming either a BEC or a non-BEC state inside a NS, depending on DM observables, namely $m_\chi$, $\sigma_{\chi n}$ and $\expval{\sigma v}$. One of the important implications of our analysis is that direct detection experiments remain relevant for the electroweak scale DM (see Fig.\ref{fig:scan}). Indirect detection of DM puts an upper limit on the $\expval{\sigma v}$, whereas our work provides a lower limit - keeping the mass around the electroweak scale unconstrained as well (see Fig.\ref{fig:sigmav}).

As is apparent that the constraints on DM parameters are sensitive to how BH forms from different thermal states of DM particles inside the NS. So, the question is whether we can observationally distinguish two kinds of BHs, namely BH originated from the BEC state and BH originated from a non-BEC state. In fact, the distinction between a BEC state BH and a non-BEC state BH hinges on the observation of the collapse time of BHs, as defined in Eq.\ref{eq:collapse}. Gravitational wave signals associated with binary systems of NSs and BHs can be used to infer about the collapse time, which is quite different for two kinds of BHs in consideration. In a binary NS population, the typical time delay to complete the merger event can be as large as  $\mathcal{O}$(Gyr)\cite{Zevin:2022dbo}, whereas the collapse time is rather small as shown in Eq.\ref{eq:colltime}. Consequently, some of binary NSs in a population can implode into binary BHs before merging into one another. Future gravitational wave detectors can measure the relative fraction of binary NSs and binary BHs in such a population, in turn inferring about the collapse time. For computational details see Ref.\cite{Singh:2022wvw}. Therefore, with sufficient  statistics of GW events in the foreseeable future, BHs with different origins in our scenario can be distinguishable. 

Now, in the absence of BH formation from the host NS, the distinction between a BEC  and a non-BEC DM core can be possible via probing the equation of state (EOS) of the NS \cite{Ellis:2018bkr,Karkevandi:2021ygv,Buras-Stubbs:2024don,Shakeri:2022dwg}. In particular, the EOS of a NS is modified differently for a BEC and a non-BEC DM core, as DM particles in the BEC state essentially are confined within a smaller spatial region with negligible pressure than the latter. This is reflected in two NS observables, namely the \textit{mass-radius relation} and the \textit{tidal deformability}, which  are measurable in the gravitational wave spectrum \cite{LIGOScientific:2017vwq,Annala:2017llu} of binary NS systems as well as in radio observations of pulsars. In a NS binary system, tidal forces would be different for a BEC and a non-BEC DM core, which would be reflected in the tidal deformability parameter. However, the exact merger dynamics is rather involved, therefore left for elsewhere \cite{ghosh}.
\section*{Acknowledgment}
We acknowledge Debajit Bose, Tanmoy Kumar, Arunava Mukherjee, Rajesh Kumble Nayak and Rohan Pramanick for various useful discussions. Special thanks to Anupam Ray and Aritra Gupta for their inputs at the initial phase of this work. DG acknowledges the Post-doctoral Fellowship from IISER Kolkata. 


\end{document}